\documentclass [aps,pra, onecolumn,14 pts,a4paper]{revtex4}
 \pdfoutput=1

\usepackage{amssymb,amsmath,amsfonts,graphicx}
\usepackage{bm}
\usepackage{epsfig,color,times}
 \usepackage{bm,color}
\usepackage{graphicx}
\usepackage{amssymb}
\usepackage{epstopdf}

\begin{document}

\title{Scaling laws in turbulence}

\author{Yves Pomeau$^1$ and Martine Le Berre$^2$}
\affiliation{ $^1$ Ladhyx (CNRS UMR 7646), Ecole Polytechnique, 91128 Palaiseau, France 
\\$^2$  Ismo (CNRS UMR 8214), Universit\'{e} de Paris-Saclay,
91405 Orsay, France}

\date{\today}

\begin{abstract}
Following the idea that dissipation in turbulence at high Reynolds number is by events singular in space-time and described by solutions of the inviscid Euler equations, we draw the conclusion that in such flows scaling laws should depend only on quantities appearing in the Euler equations. This excludes viscosity or a turbulent length as scaling parameters and constrains drastically possible analytical pictures of this limit. We focus on the law of drag by Newton for a projectile moving quickly in a fluid at rest.  Inspired by  the Newton's drag force law (proportional to the square of the speed of the moving object in the limit of large Reynolds numbers),  which is well verified  in experiments when the location of the detachment of the boundary layer is defined, we propose  an explicit relationship between Reynolds's stress in the turbulent wake and quantities depending on the velocity  field (averaged in time but depending on space), in the form  of an integro-differential equation for the  velocity which is solved   for a  Poiseuille flow in a circular pipe.
\end{abstract}

\maketitle

\section{Introduction}
\label{Introduction}
Turbulent flows at large Reynolds number display such a complex behavior in space and time that it is impossible to obtain a solution of the fluid equations representing well the flow field in such situations. This makes necessary a statistical description of the fluctuations in those flows. A major issue met when developing this statistical picture is the choice of  physical parameters that should enter in such a theory.  Kolmogorov suggested (see section 33 in \cite{LL}) to use the dissipated power per unit mass on average as a scaling parameter for the statistics of turbulence, assumed to be homogeneous and isotropic. Even though this last assumption is done very often in theoretical works, its relevance to explain physical situations is far from obvious. Specifically, most turbulent flows, if not all of them, have a geometrical structure making them non homogeneous and non isotropic. This obvious remark is behind developments presented below: assuming a time averaged velocity field depending on space, we derive out of it Reynolds's stress tensor by assuming that the relationship does not introduce quantities besides the one related to the average velocity field, a solution of Euler equations. The irreversibility in the dynamics of the Euler velocity is introduced by the occurrence of finite time singularities of this Eulerian dynamics: in such singularities some energy is dissipated by higher order effects in the Enskog expansion , but the amount of energy is given by the Euler equations. 

 Kolmogorov scaling assumption  agrees with the von Karman-Howarth relation \cite{LL} for the triple velocity correlation. It amounts to assume that, when viscosity is negligible, a velocity fluctuation $\delta u$  at a typical distance $\delta r$ varies like the cubic root of this distance, namely that 
$ \delta u \sim (\epsilon \delta r)^{1/3}$,  $\epsilon$  being the power dissipated per unit mass in the fluid. Taking the cube of both sides of this scaling relation, one gets the von Karman-Howarth relation. However this scaling law does not seem to be consistent with the Euler equations for an incompressible flow. This follows from a simple estimate of the local acceleration, this acceleration having the same order of magnitude as $u \nabla u$. Plugging into $u \nabla u$ the velocity $\delta u$ one finds an estimate of the acceleration diverging like $ \epsilon^{2/3} \delta r^{-1/3}$ as $\delta r$ tends to zero and for a given $\epsilon$. This is inconsistent with the fact that this acceleration is the time derivative of the velocity: at short times $\Delta t$ the velocity increment derived from the acceleration is of order $\Delta t$ times the time derivative of the velocity, of order $\delta r^{-1/3}$. This adds to the velocity a quantity of order $\Delta t  \epsilon^{2/3} \delta r^{-1/3}$, far bigger as $\delta r$ tends to zero than the estimate $\delta r^{1/3}$ one started from. This large fluctuation of the acceleration is more than a mathematical bug evidenced by this calculation, it is also something that is seen, although quite indirectly, in records of velocity fluctuations in a high speed wind tunnel. There, contrary to what is implied by the estimate $ \delta u \sim (\epsilon \delta r)^{1/3}$, the large fluctuations of acceleration do not happen all the time, but are strongly correlated to sparse large velocity fluctuations  \cite{YPMLB}. Such a correlation contradicts Kolmogorov scaling law: in this scaling law the parameter $\epsilon$ is just the product of the velocity and of the acceleration. Had this parameter a fixed value, or at least without too large fluctuations, large velocities should be correlated to small accelerations and conversely large accelerations correlated to small velocities, exactly the opposite of what is observed. 

Therefore it seems meaningful to reconsider the scalings laws for turbulence. Consistent with the scaling law $ \delta u \sim (\epsilon \delta r)^{1/3}$, a  possibility could be that the velocity field is a Holder-like function of the position with exponent $1/3$. This would make the velocity field continuous but not derivable, but the argument presented in the previous paragraph goes against this concept of a velocity being a Weierstrass-like function \cite{W} of the position at all times and (almost) everywhere. Another possibility for the dependence of the velocity field as a function of position and time was considered in 1934 by Leray \cite{leray}. It assumes that the solutions of the fluid equations display finite time singularities that are also localized in space at points.  Here we start from this assumption, supported by  our recent investigations of turbulent flows. 
By studying time records of the velocity obtained in the big wind tunnel of Modane we found \cite{YPMLB}-\cite{YM}, we believe, convincing evidence that the occurrence of such singularities explains well the observed intermittency of the signal. This is true for the strong correlation between large values of the velocity $ {\bf{u}}$ and of the acceleration $\gamma=\partial_{t} {\bf{u}}$,  at the same point in space-time. The explanation of this correlation relies on the link between  large fluctuations observed in the recorded signal  and the occurrence of Leray-type singularities, in the form
\begin{equation}
 {\bf{u}}( {\bf{r}}, t) = (t^*- t)^{-\alpha}  {\bf{U}} ( {\bf{r}}(t^*- t)^{\alpha-1})\textrm{,}
\label{eq:self-sim}
\end{equation}
  where $t^*$ is the time of the singularity, and  $\alpha$ is a  positive exponent  smaller than unity. We have shown  that  the values $\alpha$  equal  $1/2$  and  $3/5$ correspond respectively to self-similar solutions which conserve  the circulation  and the energy in the singular domain.
   
For such solutions  the space domain of large fluctuations shrinks with a (positive) power of  $(t^*- t)$, the time difference between the time of measurement and the time of blow-up. Moreover the amplitude of velocity and of the acceleration diverge with respect to this time difference according to power laws. 
Another element in favor of the occurrence of finite time singularities is the behavior of the structure function for the  Eulerian acceleration,
 \begin{equation}
{\mathcal{S}}_{n}(r)=< {\bf{\delta \gamma}}^{n}>
\label{eq:Sn}
\end{equation}
where $\delta{\gamma} ={\bf{\gamma}}(r+r_{0}) - {\bf{\gamma}}(r_{0})$, When the order $n$ in ${\mathcal{S}}_{n}(r)$ changes \cite{YM} the behavior of this function of $r$ changes dramatically.  At small   order the structure function built with the acceleration tends smoothly to zero as the distance  $r$   between its two arguments tends to zero whereas it diverges at small distances when the  order  is larger than a critical value.  We showed  that the  change of behavior  of ${\mathcal{S}}_{n}(r)$ (with respect to $n$)  is a straight consequence of the existence of singularities. 
 This  property results from the calculation of  ${\mathcal{S}}_{n}(r)$ as an  integral  over the space-time volume  of  ${\bf{\delta \gamma}}^{n}$ weighted by the  probability of recording a singularity at a given point of space-time, which is small at small distance. 
 We showed that the structure function itself diverges at small $r$ values with an exponent $p(n)$ depending linearly  on  $n$. When $n$ is less than a critical value the result of the integration defining the structure function tends to zero at small distances, but when the exponent of the structure function becomes bigger than a critical value, the divergence of amplitude of the singularity overcomes the small probability of observing a singularity so that the structure functions become divergent at small distances. The same non trivial behavior of the structure function is observed in a  non-linear-Schr\"{o}dinger model  equation which displays numerically turbulent solutions and quasi-singular events,  in a limit equivalent to the one of zero viscosity  \cite{NLS}. Furthermore the structure functions of acceleration recorded in wind tunnels at large distances tend to a constant because they represent contributions of quasi independent singularities.
    
     There remains the non trivial question (in fluid mechanics) of the effect of a small viscosity on those singular events. It is widely believed that at small scales of turbulence viscosity becomes dominant. In collapsing self-similar solutions the space scale tends to zero and so makes the viscous stress bigger and bigger as the singularity time is neared, but at the same time the velocity fluctuation becomes larger and larger. So it may happen that this velocity increase overcomes the increase of viscous stress due to the shrinking of length scales. This is exactly what happens in the Leray equations   in the case $\alpha=3/5$ which corresponds to finite energy conserved  in the shrinking domain until
viscosity (or Burnett coefficients) become effective.
We showed that in this case the perturbation of Euler dynamics due to viscosity diverges less strongly than the non linear term as one approaches the singularity, so that the viscous pressure becomes a negligible perturbation near the blow-up time. Therefore, at least in this case, there is no guarantee that viscosity regularizes the solution at the singularity time.  We refer the interested reader to the original publication on this question \cite{YM}, the conclusion being that dissipation takes place always near the singularity time but not necessarily because of the effect of viscosity. To summarize this discussion, the "natural" evolution of solutions of the fluid equations is toward finite time singularities where the dissipation of energy takes place,  due to the drag force exerted around singular structures. This kind of information is sufficient, as shown below, to derive an explicit and closed set of equations for the average velocity field, that finally yields a dissipation term independent of the viscosity.  
 

\section{Expression of Reynolds's stress}
\label{Parameters}
From what was concluded at the end of previous section, dissipation in highly turbulent flows takes place in Leray-like singularities of solutions of the Euler dynamical equations. Because of their atomic structure there cannot be actual singularities in continuous media like fluids and those singularities are smoothed out by some sort of correction to the Euler equations taking into account the molecular structure of the fluid. Usually the role of this regularizing effect is played by viscosity, but as just said,
 it is not necessarily true that viscosity cuts the singularity before blow-up. As shown in reference  \cite{YM} higher order terms in the Enskog expansion- in an extended form- do smoothen the singularity. The precise way they do it is not relevant here. It relies on higher order terms in the Enskog expansion, the so-called hyper viscosity Burnett terms, or more accurately the "renormalized Burnett terms", since the coefficients of those terms are given by diverging Green-Kubo integrals and so the hyper viscosity contributions to stress in the fluid equations have to be renormalized so to include non-local integrals (or if one prefers non integer Liouville space derivatives). The next step in this theory is to find what is the effect of such singularities on the global fluid dynamics averaged over time. 
 
\subsection{ Reynolds's stress due to the drag.}
 Here we investigate the role of the drag force around singular events, which could be a relevant source of dissipation.
At the very beginning of fluid mechanics, as we know it, was the law of drag by Newton for fast moving projectiles in air. By a very clever reasoning Newton showed that the drag force on an object moving at speed $U$ in a gas at rest at infinity is of order 
 \begin{equation}
F^{(N)} = \frac{1}{2} C_x  \rho U^2 S , 
\label{eq:FN}
\end{equation}
where  $\rho$ is the mass density of the fluid and $S$ area of the cross section of the moving object in the direction perpendicular to the velocity. Lastly $C_x$ is a numerical coefficient which, to Newton's surprise, depends on the shape of the object in a complex way that he was unable to explain. As well known too, d'Alembert showed that in a perfect inviscid fluid there is a steady flow around a moving object exerting no drag on it (see sections 9 and 11 in ref  \cite{LL}). In the familiar example of a sphere moving at constant speed, Euler's equations can be solved explicitly and d'Alembert paradox applies so that no drag should be exerted on the sphere. Neither Newton nor d'Alembert refer to turbulence, a concept unknown at their time. The resolution of d'Alembert paradox is that dissipation takes place in the wake of the moving object and yields ultimately a drag in agreement with Newton's insight. In retrospect Newton's law for the drag is an early version of what is called sometimes the zeroth law of turbulence, namely that, in the limit of high Reynolds number, dissipation tends to a value that does not depend on viscosity (this amounts to say that, in the limit of large Reynolds number, Newton's $C_x$ coefficient tends to a constant, something in full agreement with observations  \cite{GF}). 

The connection between the drag law of Newton and our search of scaling laws for fully developed turbulence comes from the fact that, physically, Newton's drag represents the dissipation by turbulence in the wake of a fast moving body. Therefore the scaling laws for the representation of this turbulent wake should be consistent with the final result, namely that the drag is independent on the viscosity. Therefore this drag must depend only on quantities occurring in the inviscid equations. This is fully consistent with the idea of dissipation by finite time singularities, because the amount of energy in such a singularity is independent on viscous phenomena and is dissipated at the singularity time. This leaves few parameters  to model the turbulent flow responsible of the drag. As we are looking for time-averaged quantities one needs to look at equations for averages only, without time dependence. Compared to the standard fluid equations, there is no obvious constraint imposing that the equations we are looking for are local in space, namely written with space derivatives of finite order at a given point. This allows to write equations with integral terms. Let us look at the possibility of a new term with an integral kernel and the same physical dimension as the $u \nabla u$ term in Euler equation for momentum. 
Because this term should be the gradient of Reynolds's turbulent stress, this stress has to be quadratic with respect to the velocity (see section $15$ in Ref.\cite{LL}). 

To agree with the order of magnitudes derived from the idea of dissipation only due to processes described by Euler equations, let us come back to the expression of Newton's drag on an object moving at speed $U$ in a fluid at rest at infinity. The $i-$ Cartesian component of this drag force $F^{(N)}$ and is given by 
 \begin{equation}
F_i^{(N)} = - \frac{1}{2} C_x  \rho S U_i  \vert U \vert,
\label{eq:FiN}
\end{equation}
where $\vert U \vert$ is the modulus of the vector $U$. Such a law, because it is non analytic with respect to the velocity, cannot be derived from a theory of Reynolds's stress that is analytic with respect to the average velocity field and its space derivatives. Moreover we aim at deriving such a law of stress that satisfies standard requirements of invariance with respect to rotation of the coordinate system, and-what is far more difficult- by respecting the same scaling as the Euler equations, namely without introducing  parameter other than in the velocity field itself. This requirement can be satisfied by choosing a law for Reynolds's stress that depends on the modulus $\vert U \vert$. 

It is worth pointing out that equation (\ref{eq:FiN}) is not the most general for Newton's drag on an object in a fast flow. It applies only for objects with a high degree of symmetry like a sphere. If one takes a less symmetric object, without downstream/upstream symmetry, the drag is a priori different when the speed changes sign. This can be accounted for by adding to the expression a term analytical wit respect to $U$, namely of the form $D_{ijk} U_j U_k$ where $D_{ijk}$ is a rank three tensor depending on the shape of the object and non-zero if the object is not symmetrical with respect to exchange of opposite directions of space. Moreover for such non symmetric objects one has to substitute for the constant $C_x$ in equation (\ref{eq:FiN}) a rank two tensor $C_{ij}$, so that the general expression for Newton's drag law becomes 
\begin{equation}
F_i^{(N)} = - \frac{ \rho}{2} S (C_{ij} U_j  \vert U \vert + D_{ijk} U_j U_k)
\label{eq:FiN.1}
\end{equation}
where $S$ now is an the area of an arbitrary cross section and summation on identical indices is implied. The choice of this cross section changes of course the values of the dimensionless tensors $C_{ij}$ and $D_{ijk}$. Moreover, because there is always friction, the drag must be such that the scalar product $F_i^{(N)} U_i$ is always negative for any orientation of the velocity $U_i$.

To get some insight on possible laws for this stress, let us look at the standard viscous stress  tensor. It is given by 
 \begin{equation}
\sigma_{ij}^{v} = - \eta (u_{i, j} + u_{j,i}) ,
\label{eq:stressvisq}
\end{equation}
where $u_{i, j} = \frac{\partial u_i}{ \partial x_j}$,  $i, j, ...$ being the indices of Cartesian coordinates and $ \eta $ the shear viscosity of the fluid. Reynolds's stress tensor should have the same physical dimension as the inertial stress $u_i u_j$ (assuming both to be multiplied by the mass density $ \rho$, not written explicitly later on). Therefore, starting from the expression of the viscous stress to transform it into the one valid for Reynolds's turbulent stress one has to multiply it by $ \vert u\vert$, $u$ being now understood as the local  \textit{time-averaged velocity} with respect to a boundary (a plane in a Poiseuille-like geometry or the surface of a moving object in a wake problem). This yields an expression of the turbulent stress depending on the mean (over time)  velocity like  $\vert u \vert (u_{i, j} + u_{j,i})$. The physical dimension of this quantity with respect to the velocity is, as desired, quadratic with respect to the velocity. However, supposing the dimension of velocity fixed,  the dependence with respect to length scale $L$ is like $1/L$, because of the derivative of the velocity with respect to the position variable. This has to be changed because one does not want to introduce in the value of Reynolds's stress any quantity depending explicitly on a length defined independently on the velocity field.  

Another remark relevant for the existence of the absolute value $\vert u\vert $ in
the above (partial) expression for the stress tensor comes from an
analogy with Maxwell expression for the shear viscosity of gases. This
shear viscosity is proportional to a quantity with the physical
dimension  $\vert u\vert \ell $  where   $\vert u\vert $  is the  typical thermal speed
of a particle, and $\ell$ the typical mean free path. In Maxwell kinetic theory of viscosity the mean free
path is the distance over which the momentum difference between two
particles is carried, ending with the annihilation of this momentum
difference by their collision. In a singular Leray-like event, an
initial difference of momentum between  parts of fluid
participating to the collapse is cancelled when the velocity field of
the domain shrinks to zero, which erases the initial momentum
difference, and so participate to the transfer of momentum in space.   The role of those parts of the fluid with different velocities is played in Maxwell's theory of  a dilute gas, by the two particles on their path towards a collision (which explains ultimately the shear viscosity of the dilute gas).


To somehow "multiply" the combination $\vert u \vert (u_{i, j} + u_{j,i})$ by a length one has to integrate it with respect to the position variable but by keeping the requested independence of the result under rotation of the system of coordinates. This excludes for instance to carry an integration with respect to only one space variable because the choice of one coordinate obviously breaks the symmetry under rotation of the system of coordinates. The only convenient expression is found by integrating $\vert u \vert (u_{i, j} + u_{j,i})$ computed at $X'$ (capital letters being for position in space, {$X=(x,y,z)$), over the whole space with the integrand $\frac{1}{\vert X - X' \vert^2 }$, the power of the denominator being chosen to give the whole transformation the scaling of a length.  The final result is 
    \begin{equation}
  \sigma_{ij}^{Re}(X) = K  \int {\mathrm{d}} X' \frac{ \vert u (X')  \vert (u_{i, j} + u_{j,i})(X')}{ \vert X - X' \vert^2} 
  \textrm{,}
\label{eq:aver1}
\end{equation}
This is the equation for Reynolds's stress compatible with the requirement that no parameter other than what appears in Euler equation is introduced in the phenomenological relation between  the average velocity field and Reynolds's stress. The quantity denoted as $K$ is a pure number, independent on the chosen system of units. 

It is of interest at this point to explain the connection between the law for Reynolds's stress in equation (\ref{eq:aver1}) and Euler equations. As we said this law is non analytic with respect to the velocity field because of the absolute value $ \vert u (X')  \vert $ appearing in it. This shows also that it cannot be derived from Euler equation by a calculation using standard algebra. The physics of this law is that finite-time singularities dissipate energy and so introduce a fundamental irreversibility in solutions of equations (the Euler equations) which are formally reversible with respect to time. Deriving irreversible equations from reversible dynamics is fairly standard in theoretical physics, the most famous example being the derivation of Boltzmann kinetic theory from the reversible Newtonian dynamics of particles. In the present case, the irreversibility is a consequence of the dissipation taking place at the time of the singularity of the self-similar solution of Leray-like equations. As shown below in the case of Poiseuille flow, the irreversibility of the equation is tightly  linked to the absolute value in the equation for the turbulent stress  because thanks to it, the average flow in a pipe is reversed as  the pressure gradient is reversed. The absolute value is a consequence, although slightly hidden, of the existence of dissipation in singular events. Otherwise  (if  the equation for the turbulent stress doesn't contain absolute value of the velocity)  there is no dissipation by the Euler equation.
 This explains, hopefully, why equation   (\ref{eq:aver1}) cannot be derived directly from Euler equation by assuming their solution to be smooth. An added information is necessary, namely that solutions of Euler equations have finite time singularities where energy is dissipated. 

\subsection{ Poiseuille flow in a pipe} 
\label{sec:Pois}
An example showing how to use this equation for specific problems of fluid mechanics is the case of Poiseuille flow in a pipe with a cross section left arbitrary for the moment, this pipe being alined along the $x$- direction. Let us consider for this simple case the calculation of the integral in equation (\ref{eq:aver1}). For that purpose we define coordinates  in the three dimensions of space as $x$ in the direction of the pressure gradient driving the fluid, $y$ and $z$ being the coordinates perpendicular to $x$. 
We shall use the standard notation for the components of the velocity, $u$ in the $x$-direction, $w$ in the $z$- direction and $v$ in the $y$-direction. Let us assume first that the velocity field has the same geometry as in the Poiseuille case, namely that only the component $u$ is not zero and that it depends on $(y, z)$ only. 
Therefore the only non-vanishing components of   $\sigma_{ij}^{Re}$ are 
 \begin{equation} 
   \sigma_{xy}^{Re}(y, z) = K  \int {\mathrm{d}} z' \int  {\mathrm{d}} y'  \int  {\mathrm{d}} x' \frac{\vert u \vert u_{, y}(y', z')}{(z- z')^2 + (y - y')^2 + x'^2}  . 
   \label{eq:sigxy}
\end{equation}   
and another similar equation for $\sigma_{xz}^{Re}(y, z)$. From the symmetry of the stress tensor, 
one has $\sigma_{xy}^{Re} = \sigma_{yx}^{Re}$ and $\sigma_{xz}^{Re} = \sigma_{zx}^{Re}$ for the two other non-zero components.
 
In the case of a plane Poiseuille flow, there is a priori no dependence with respect to the variable in the $y$-direction (if $y$ is the coordinate along the planes limiting the fluid and perpendicular to the average velocity) and the integral diverges logarithmically at large distances in the $(y', z')$ plane.  This was to be expected because the theory under consideration  has no internal length scale like Prandtl mixing length for instance. Therefore, in this case, the interactions somehow can be cut only by a length related to geometrical conditions on the boundaries. In the present case this amounts to say that the turbulent Poiseuille solution is more complicated than what was imagined first and that the solution with an infinitely extended geometry in the $y$-direction is not independent on the $y$ coordinate. It implies that either this solution is dependent on this $y$-coordinate in a non-trivial way, for instance periodically, with a wavelength proportional to the vertical distance between the two plates, or that the problem makes sense only for a finite-but large-width in the $y$-direction.  

Let us look at a simpler situation, namely the one of a Poiseuille flow in a pipe with circular cross section of radius $R$. In this case, the various quantities involved depend on the radial distance only, namely $r = (y^2 + z^2)^{1/2}$. This gives the only non vanishing component of Reynolds's stress (written with the cylindrical coordinates $(x, r)$):  
 \begin{equation} 
   \sigma_{xr}^{Re}(r) =  K  \int_0^R {\mathrm{d}} r' r'  \vert u \vert u_{,r}(r')  \int_{- \infty}^{+ \infty} {\mathrm{d}} x' \int_0^{2  \pi}  {\mathrm{d}}  \varphi \frac{1}{r^2 + r'^2 - 2 r r'  \cos( \varphi) + x'^2}
\label{eq:sig1}
\end{equation}   
This yields after integration on the variable $x'$: 
 \begin{equation} 
   \sigma_{xr}^{Re}(r)= 2  \pi K  \int_0^R {\mathrm{d}} r' r'  \vert u(r')  \vert u_{,r}(r') \int_0^{2  \pi}  {\mathrm{d}}  \varphi \frac{1}{(r^2 + r'^2 - 2 r r'  \cos( \varphi))^{1/2}}
  \label{eq:sig2}
\end{equation}    
 The result of the integration on the angle $ \varphi$  can be expressed by means of  complete elliptic integrals 
 of the first kind 
 \begin{equation} 
 I(k )= \int_0^{ \pi/2}  {\mathrm{d}}  \theta	\frac{1}{(1-k^{2}\sin^{2} \theta )^{1/2} }
 \textrm{,}
\label{eq:ellip-integ}
\end{equation} 

 In (\ref{eq:sig2}),  the integration over $\varphi$  for  $0<\varphi<2\pi$ is twice  the integration over $0<\varphi<\pi$ . Setting  $k_{-} = 2i \sqrt{rr'}/ \vert r-r' \vert$ , the integration over $\varphi$  for  $0<\varphi<\pi$, gives $\frac{ 2I( k_{-}) }{\vert r-r' \vert} $, and setting
$k=2 \sqrt{rr'}(r+r') $,  the integration over $\varphi$  for  $\pi<\varphi<2\pi$, gives $\frac{ 2I( k) }{ r+r'} $.  Using the latter result,  the Reynolds stress in a Poiseuille flow along a pipe is given by the integral
 \begin{equation} 
   \sigma_{xr}^{Re}(r)= 8 \pi K  \int_0^R {\mathrm{d}} r'  r' \vert u(r')  \vert u_{,r'}(r') \frac{ I( k) }{ r+r'} 
\textrm{,}
\label{eq:aver1.4.1}
\end{equation}

Reynolds's stress is to be added  to the  term $u_{i} u_{j}$   in the   equation  for the time-stationary stress tensor, deduced from the Euler equation for the momentum, that becomes with the present formalism 

 \begin{equation}
   \partial_i (u_i u_j +\sigma_{ij}^{(Re)}) +   \partial_j p   =  0 
 \textrm{,}
\label{eq:aver1.4}
\end{equation}
The condition of incompressibility $ \frac{\partial u_i}{ \partial x_i}= 0$ is to be added to this equation. The boundary condition for the unknown $x$- component of the averaged velocity field $u(r)$ makes a non trivial issue, because Reynolds's stress depends on this field in a non local way. From the way this stress was built one may guess that the boundary condition is the same as for a viscous fluid, namely that the velocity on the surface of the solid is the local velocity of this solid, which amounts to impose $u = 0$ at the radius $r = R$ of the pipe. To take a physical case where the boundary condition on the bounding surface plays an important role, we recall the experimental work by Nikuradse  \cite{nik} in the nineteen thirties who reported an significant sensitivity of the turbulent friction with respect to a controlled roughness of the pipe wall. This is hard to explain if one assumes that turbulence is generated by large scale motion inflows decaying to small dissipative scales by non linear interaction. 

In the theory presented here this sensitivity to surface roughness could follow from the substitution to the regular boundary condition of continuity of tangential velocity along a smooth wall by a so-called mixed condition of the Navier type like 
 \begin{equation}
  u_t +  \lambda  \frac{ \partial u_t }{\partial n} = 0  
   \textrm{,}
\label{eq:ut}
\end{equation}
 where $u_t$ is the component of $u$ tangent to the surface and $n$ the coordinate locally  normal to this surface. The quantity denoted as  $ \lambda$ has the physical dimension of a length. It represents, for a rough surface the root of the mean square deviation of the fluctuations of height of the rough  surface. With such a boundary condition, the solution of equation (\ref{eq:aver1.5}) below for Reynolds(s stress depends on $ \lambda$. Because there is no length parameter in the problem besides geometrical parameter of the flow, like the diameter of pipes in Poiseuille flows, it could be that the final result, namely the drag per unit length of the pipe, depends sensitively on the length $\lambda$, as observed. 
This contrasts with the predictions of a theory based on the idea that large scale motion dominates dissipation by transfer of energy toward small scales: if one follows this idea it seems very hard to understand the sensitivity of observed dissipation to roughness of the boundaries, a roughness that  changes small scales of the Euler dynamics only. 

The explicit  form of equation (\ref{eq:aver1.4}) to be solved for a circular pipe submitted to an uniform pressure gradient in the $x$-direction is 
\begin{equation}
 \frac{1}{r} \frac{\partial (r \sigma_{xr}^{Re}(r))}{\partial r} + C = 0
\textrm{,}
\label{eq:aver1.5}
\end{equation}
where $C = p_{,x}$ is a constant, which will be taken positive in the following,  and where the function $ \sigma_{xr}^{Re}(r)$ is given by equation (\ref{eq:aver1.4.1}). A solution  of (\ref{eq:aver1.5}) is
\begin{equation}
  \sigma_{xr}^{Re}(r)= - \frac{C}{2}r + \frac{A}{r}
\textrm{,}
\label{eq:solsig}
\end{equation}
where A is a constant .  On the axis of the pipe  this solution diverges except  for A=0.  Finally in order to get an expression for the  time-averaged  velocity field from the quadratic term 
\begin{equation}
 \mathcal{M}(r)=\vert u(r)  \vert u_{,r}(r),
\label{eq:M}
\end{equation}
  one has to introduce the solution $  \sigma_{xr}^{Re}(r)= - \frac{C}{2}r$ into (\ref{eq:aver1.4.1}). Setting $z=r/R$, $z'=r'/R$, 
\begin{equation}
\tilde{u}= \frac{u}{c^{1/2} }    \qquad   c = \frac{C}{16 \pi K},
\label{eq:tildu}
\end{equation}
 and  defining $ \tilde{\mathcal{M}}(z)=\vert \tilde{u}  \vert \tilde{u}_{,z}(z)$ which is equal to  $\mathcal{M}/c^{2}$ , one get the  following scaled integral equation for $\tilde{\mathcal{M}}(z)$  
\begin{equation}
  \int_0^1 {\mathrm{d}} z'  \tilde{ \mathcal{M}}(z') \frac{x'}{ z+z'} I[k(z',z]  = \;- \;z,
\label{eq:inteq}
\end{equation}
where $k(z',z)=2 \sqrt{zz'}(z+z')$. Note that using  the above scaled   quantities,  (\ref{eq:inteq} )  is parameterless. 
A numerical solution of  (\ref{eq:inteq} )  is obtained by approximating the integral in the l.h.s.   as a Riemann sum where the integrand is the  product of a  matrix $m$ by a vector $v$, the matrix  elements of $m$  and $v$  being  respectively the discrete values of $ \frac{z'}{ z+z'} I[k(z',z)]$ and  of $\tilde{ \mathcal{M}}(z')$. Then inverting the matrix $m$, the solution  for the vector $\tilde{ \mathcal{M}}(z_{i})$, $0 < z_{i} <1$,  written as
\begin{equation}
  \tilde{ \mathcal{M}}(z_{i})= m^{-1}(-z_{i}),
\label{eq:sol-inteq}
\end{equation}
  is shown in Fig. \ref{fig:Mu}.  We note that the curve changes sign midway between the pipe axis, the positive and negative parts being dissymmetric, with   a strong decrease close to the boundary.
The mean velocity $u(r)$ along a diameter can be deduced as follows. First we may set $\vert u\vert =u$, or $M(r)=u u_{,r}$, because   the mean velocity keeps the same  sign in the plane perpendicular to the $x$ axis  of the pipe, its sign being the one of the pressure gradient, or $C$, assumed positive  above.  Then, introducing $  \tilde{ \mathcal{P}}(z) = \int_0^z {\mathrm{d}} z'   {\mathrm{d}} z'  \tilde{ \mathcal{M}}(z')$ ,  we get the relation
\begin{equation}
  \tilde{ \mathcal{P}}(z)=  ( \tilde{u}^{2}(z) - \tilde{u}^{2}(0))/2.
\label{eq:P}
\end{equation}
The on-axis and boundary values of the velocity are linked  by two relations. First (\ref{eq:P}) gives  $\tilde{u}^{2} (1)= \tilde{u}^{2} (0)+ 2\tilde{ \mathcal{P}}(1)$. Moreover 
 using (\ref{eq:ut}), one has  in scaled variables $ \tilde{u}^{2} (1)= -\tilde{ \mathcal{M}}(1) \lambda/R$. In summary  the scaled on-axis value of the mean velocity $ \tilde{u}(0)$  depends on $\lambda/R$ only and can be deduced from the boundary values of two numerical curves displayed in fig.\ref{fig:Mu}. In physical variables $u(0)$  depends additionally on the constants $C,K$ via the relation
\begin{equation}
u(0)^{2}= - \frac{C}{16\pi K} (\tilde{ \mathcal{M}}(1) \lambda/R + 2\tilde{ \mathcal{P}}(1) ),
\label{eq:uo}
\end{equation} 
which agrees with the observations that in the turbulent regime  when the roughness decreases, the  losses decreases.

    \begin{figure}
\centerline{ 
(a)\includegraphics[height=1.5in]{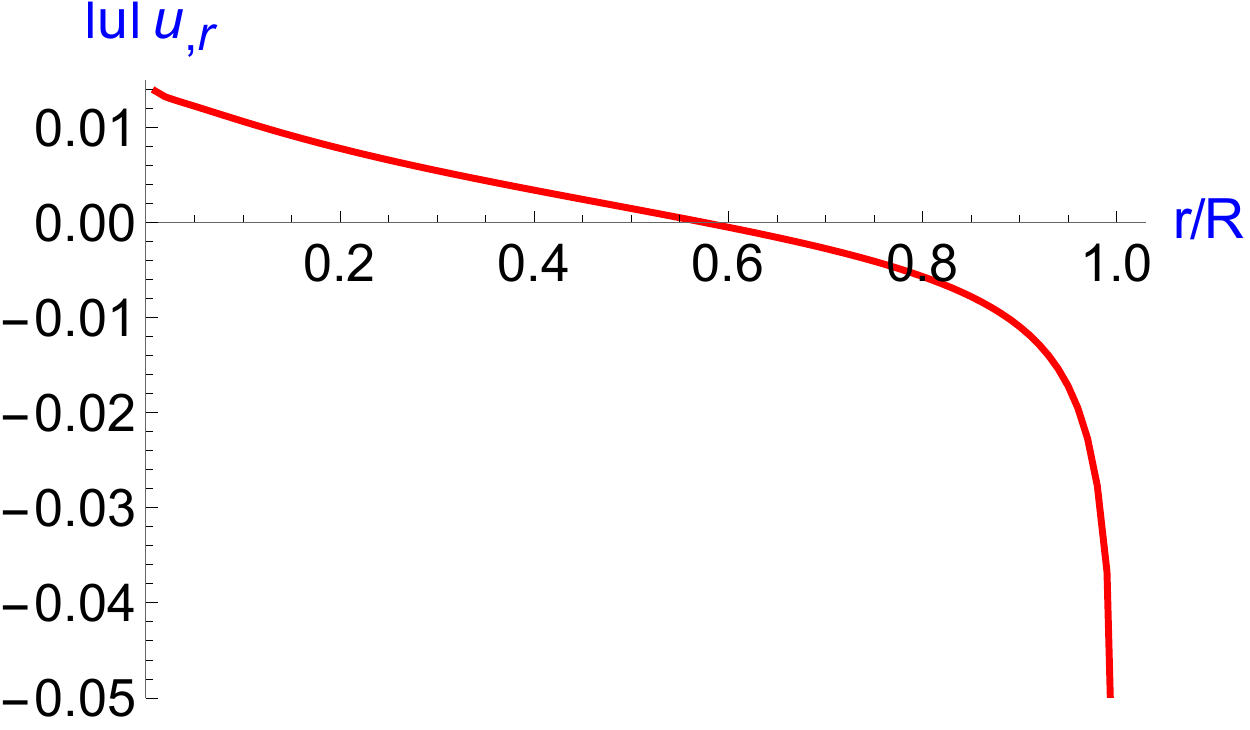}
(b)\includegraphics[height=1.5in]{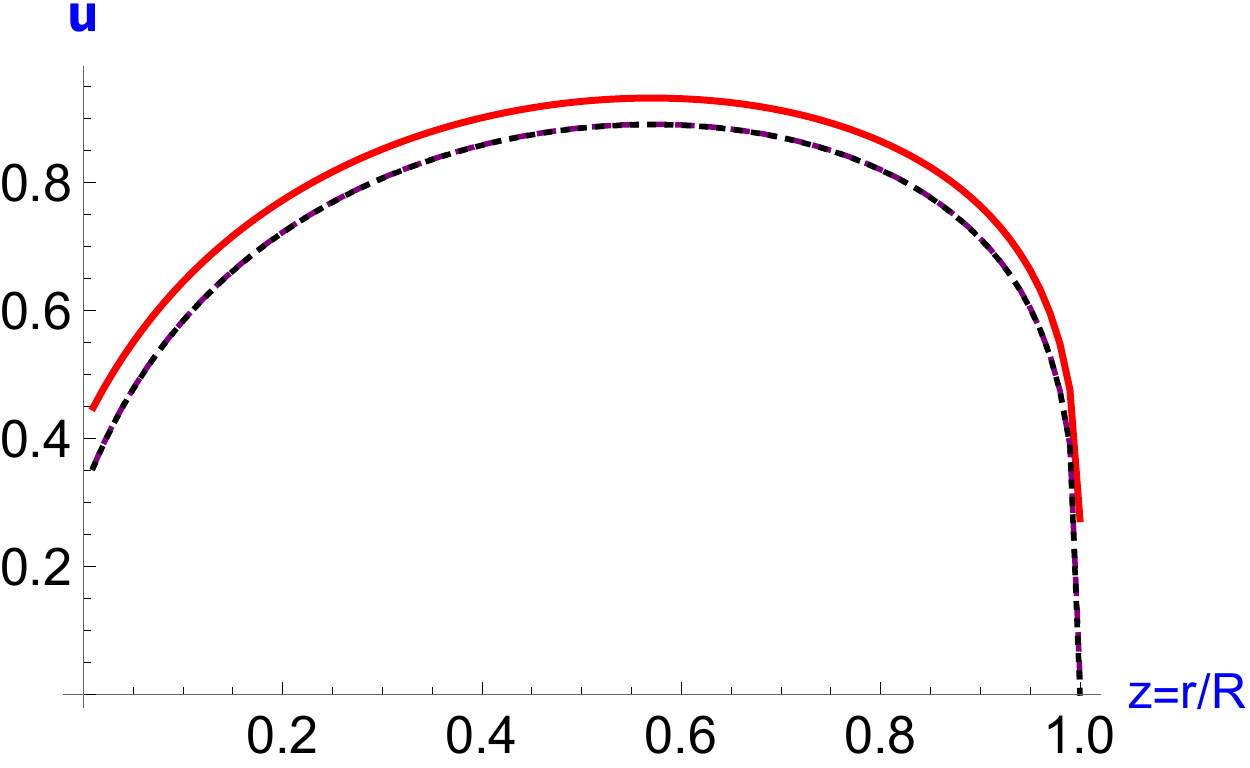}
  }
\caption{ (a) Numerical solution of  (\ref{eq:inteq} ) for $ \tilde{ \mathcal{M}} =\vert \tilde{u}  \vert \tilde{u}_{,z}(z)$ as function of  the scaled radius of the pipe, $z=r/R$. (b)  Profiles of the scaled velocity  $  \tilde{u} (z)=\sqrt{\tilde{u}^{2}(0) + 2\tilde{ \mathcal{P}}(z)} $ along a radius of the pipe.  the $3$  curves  from top to bottom   are for decreasing values of the roughness values of the roughness, $\lambda/R= 10^{-2}$ (red solid line), $10^{-4}$ (purple-dashed), $10^{-6}$ (blue dotted). The last two curves are hardly distinguishable,  except close to the boundary where  the value of $\tilde{u}(1)= u(R)/c^{1/2}$   tends  clearly to zero as $\lambda/R$ tends to zero  (clear on the numerics but unclear in the figure), although remaining positive. }
\label{fig:Mu}
\end{figure}

Let us point out that, contrary to the viscous case, all terms in the equation (\ref{eq:aver1.4})  for the balance of longitudinal momentum are quadratic with respect to the velocity.  It is also important to point out that the mean velocity 
in  equation (\ref{eq:aver1.4}) changes sign, as it should, when the pressure gradient driving the flow changes sign. In the case of a parallel flow in a pipe,  the inertial stress $u_i u_j$ doesn't contribute  to equation (\ref{eq:aver1.4}). By changing the sign of the velocity, Reynolds's stress $\sigma_{ij}^{(Re)}$ changes sign.

\section{Conclusion and perspectives}
\label{Parameters.1}
This work intended to explain how the centuries-old law of drag by Newton can be used to illustrate  our hypothesis that
 the dissipation occurs in fully developed turbulence by singular events,  which are local in space and time and described by solutions of the 3D Euler equations for incompressible fluids. A consequence of that is an explicit equation for the space dependent (but time averaged)  velocity, resulting from an expression of Reynolds's stress as a function of this average fluid velocity. This opens new perspectives. To take an example, the equation for the average velocity has a fairly nontrivial structure, being non linear, non local and non analytic.  Therefore it could well be that, with given boundary conditions, it has more than one solution  (a property not found in the exemple of Sec.\ref{sec:Pois}), as has its simpler counterpart for viscous fluid equations. This would fit well with the observation that turbulent flows  may show bifurcations as reported by Coles \cite{coles} in the case of circular Couette flows.

\section*{Acknowledgements}
The author thanks Christophe Josserand and Sergio Rica for useful discussions on the matter of this note. 


\end{document}